\documentclass[twocolumn]{article}
\usepackage{geometry}
\usepackage{times}
\usepackage{soul}
\usepackage{url}
\usepackage[hidelinks]{hyperref}
\usepackage{graphicx}
\usepackage{subcaption}
\usepackage{amsmath}
\usepackage{amsthm}
\usepackage{booktabs}
\usepackage{algorithm}
\usepackage{algorithmic}
\usepackage{amssymb}
\usepackage{multirow}
\usepackage{authblk}
\usepackage{epstopdf}
\usepackage{cite}
\usepackage{color}
\usepackage{tikz}
\usepackage{cleveref}
\crefname{section}{§}{§§}
\Crefname{section}{§}{§§}

\usepackage{xspace}
\usepackage{enumitem}
\usepackage{bbding}

\newcommand{\codename}{\mbox{\textsc{AlphaFuzz}}\xspace}
\newcommand{\codenameplusplus}{\mbox{\textsc{AlphaFuzz++}}\xspace}
\newcommand{\AFL}{{\textrm{AFL}}\xspace}
\newcommand{\AFLFast}{{\textrm{AFLFast}}\xspace}
\newcommand{\FairFuzz}{{\textrm{FairFuzz}}\xspace}
\newcommand{\EcoFuzz}{{\textrm{EcoFuzz}}\xspace}
\newcommand{\myparagraph}{\vspace{5pt}\noindent\textbf}
\newcommand{\ignore}[1]{}

\begin{document}

\title{Evolutionary Mutation-based Fuzzing as Monte Carlo Tree Search}

\author[1]{Yiru Zhao}
\author[1]{Xiaoke Wang}
\author[1,2]{Lei Zhao \thanks{Corresponding author: leizhao@whu.edu.cn}}
\author[3]{Yueqiang Cheng}
\author[4]{Heng Yin}
\affil[1]{School of Cyber Science and Engineering, Wuhan University, China}
\affil[2]{Key Laboratory of Aerospace Information Security and Trusted Computing, Ministry of Education, China}
\affil[3]{NIO Security Research}
\affil[4]{UC Riverside}
\date{}

\maketitle

\begin{abstract}
\small
Coverage-based greybox fuzzing (CGF) has been approved to be effective in finding security vulnerabilities.
Seed scheduling, the process of selecting an input as the seed from the seed pool for the next fuzzing iteration, plays a central role in CGF. 
Although numerous seed scheduling strategies have been proposed, most of them treat these seeds independently and do not explicitly consider the relationships among the seeds.

In this study, we make a key observation that the relationships among seeds are valuable for seed scheduling. We design and propose a ``seed mutation tree'' by investigating and leveraging the mutation relationships among seeds. With the ``seed mutation tree'', we further model the seed scheduling problem as a Monte-Carlo Tree Search (MCTS) problem. That is, we select the next seed for fuzzing by walking this ``seed mutation tree'' through an optimal path, based on the estimation of MCTS. 
We implement two prototypes, \codename on top of AFL and \codenameplusplus on top of AFL++. The evaluation results on three datasets (the UniFuzz dataset, the CGC binaries, and 12 real-world binaries) show that \codename and \codenameplusplus outperform state-of-the-art fuzzers with higher code coverage and more discovered vulnerabilities. In particular, \codename discovers 3 new vulnerabilities with CVEs.
\end{abstract}

\section{Introduction}
Coverage-based greybox fuzzing (CGF) has been approved to be very effective in finding security vulnerabilities in real-world applications and has been widely studied and used in both academia and industry~\cite{AFL,serebryany2015libfuzzer,aflfast,fairfuzz,ecofuzz,greyone}.
In general, CGF starts with several initial inputs (a.k.a. seed inputs).
Then, CGF leverages a seed scheduling strategy to select an input as the seed, and generates new inputs by randomly mutating the seed. With lightweight program instrumentation, CGF executes a program with these newly generated inputs and collects code coverage to further guide the next cycle of seed scheduling. 
In this way, CGF makes progress in discovering vulnerabilities through iterative seed scheduling, mutation, and program states exploration.

Seed scheduling, the process of selecting a seed from the seed pool for the next fuzzing iteration, plays a central role in CGF.
First, CGF utilizes a seed scheduling strategy to bridge the gap between limited fuzzing effort and an ever-increasing number of seeds.
More importantly, seed scheduling determines the directions of exploring program states. 
Specifically, CGF explores program states by dynamic execution using the inputs mutated from seeds. 
As mutations are often slight (e.g., bit-flip or byte-flip in \AFL~\cite{AFL}), the newly generated input often differs from the seed only in small input regions. 
Consequently, the execution using the new input could result in a transition to a path that is a neighbor of the path corresponding to the seed. 
In other words, dynamic executions using mutated inputs can be regarded as the process of exploring the neighbor paths of the corresponding path of the seed.
Therefore, seed scheduling refers to the selection of an execution path for exploring its neighbor paths, which determines the directions of exploring program states.

Numerous seed scheduling strategies have been proposed. Generally speaking, these strategies aim to assign a score to each seed based on certain criteria and then choose the seed with the highest score for the next fuzzing iteration. For example, \AFL~\cite{AFL} prefers seeds with the smallest size and shortest execution time. \AFLFast~\cite{aflfast} assigns more energy to the seed with low frequency. \FairFuzz~\cite{fairfuzz} prefers to choose a seed that covers more rare branches. \EcoFuzz~\cite{ecofuzz} favors newly generated seeds. These strategies treat these seeds independently and do not explicitly consider the relationships among the seeds. 

In this paper, we make a key observation that the relationships among seeds are valuable for seed scheduling, especially the seed mutation relationship (e.g., ``$A \rightarrow B$'' means Seed $B$ is mutated from Seed $A$). More specifically, a tree structure can be formed for these seeds by following their mutation relationships, which we refer to as ``seed mutation tree''. Then we model the seed scheduling problem as a Monte-Carlo Tree Search (MCTS) problem. That is, we select the next seed for fuzzing by walking this seed mutation tree through an optimal path, based on the estimation of MCTS. 

The main advantage of the ``seed mutation tree'' is that it benefits the seed scheduling to balance between exploitation  (i.e.,repeatedly exercising a high-potential seed and its neighbors) and exploration (i.e., trying a rarely-exercised seed) for optimal performance.
As demonstrated above, seed scheduling refers to the selection of an execution path for exploring its neighbor paths. With the relationships among seeds, we can further organize paths corresponding to every seed as a tree structure, which is denoted as execution tree~\cite{s2e}. In this way, we can model the fuzzing process of exploring program states as the growth of the execution tree. Based on this observation, we can leverage the ``seed mutation tree'' to approximate the execution tree, in which each path corresponds to a seed. For example, the mutation relationship (``$A \rightarrow B$'') indicates that the CGF covers a new path corresponding to $B$ by converting the result of a conditional jump on the path of $A$. As an approximation to the execution tree, the ``seed mutation tree'' enables the seed scheduling to balance exploitation (the tree's depth) and exploration (the tree's width), which indicates the fuzzing progress within a specific direction and multiple directions, respectively.

Moreover, the construction of ``seed mutation tree'' is lightweight, because mutation relationships are readily available in CGF and do not require extra instrumentation and expensive computation. 

A recent work AFL-HIER~\cite{aflhier} also leverages a tree structure to manage all the seeds and uses the UCB algorithm to perform seed scheduling. 
Our approach differs from AFL-HIER in three aspects. 
First, the insights are different. 
AFL-HIER defines different coverage sensitivity metrics and clusters seeds to reduce the impact of similarity. 
The core of our approach is to construct a seed tree using the mutation relationships, which is a lightweight and practical approximation to the execution tree.
Second, the tree structures are different.
In AFL-HIER, every internal node means a cluster of seeds that have the same coarse-grained coverage measurement. 
Every leaf node represents one seed. 
In our approach, every node in the tree refers to one seed, and the edge represents the mutation relationship between seeds. 
Third, the performances of seed scheduling are different. 
As AFL-HIER requires maintaining multiple coverage metrics for clustering, it introduces overhead to seed scheduling and finally results in a negative impact on the fuzzing throughput. 
By contrast, seed scheduling on our seed tree is lightweight. 
As throughput is a significant factor for the fuzzing performance, our approach can outperform AFL-HIER.

We implement two prototypes of the MCTS-based seed scheduling, \codename on top of AFL~\cite{AFL}, and \codenameplusplus on top of AFL++\cite{AFLplusplus-Woot20}. To demonstrate the effectiveness of our approach, we conduct a comprehensive evaluation on three datasets, including UniFuzz~\cite{UniFuzz}, CGC binaries~\cite{DARPA}, and 12 real-world binaries.
We compare the performance of \codename with AFL-based techniques \AFL, \AFLFast, \FairFuzz, and \EcoFuzz. For AFL++-based techniques, we compare \codenameplusplus with AFL++ and AFL++-HIER.
Evaluation results show that \codename and \codenameplusplus outperform other baseline techniques in terms of code coverage and discover more vulnerabilities.
Specifically, \codename achieves higher code coverage on UniFuzz than \AFL, \AFLFast, \FairFuzz, and \EcoFuzz on 14, 15, 14 and 16 out of 18 binaries, respectively. \codename and \codenameplusplus discover more unique bugs and exploitable vulnerabilities than others.
For CGC binaries, \codename discovers 87 vulnerabilities, whereas \AFL, \AFLFast, \EcoFuzz, and \FairFuzz discover 83, 83, 73, and 76 respectively. Moreover, \codename and \codenameplusplus can find bugs faster than other fuzzing techniques.
In addition, \codename discovers 3 new CVEs on 12-real-world binaries dataset.

The contributions of this study are as follows:
\vspace{-\topsep}
\begin{itemize}
\item \textbf{\emph{New insight}}. We make a key observation that the relationships among seeds are valuable for seed scheduling. We investigate and leverage the seed mutation relationships to construct a ``seed mutation tree'', which is an approximation of the execution tree for the fuzzing progress and can further benefit the seed scheduling to balance exploitation and exploration.

\item \textbf{\emph{New fuzzing technology}}. With the ``seed mutation tree'', we model the seed scheduling problem as a Monte-Carlo Tree Search (MCTS) problem and propose an MCTS-based seed scheduling strategy. This strategy strikes a balance between exploitation and exploration, due to the nature of the MCTS algorithm.

\item \textbf{\emph{Open-source implementation}}. We implement two prototypes, \codename and \codenameplusplus, and conduct comprehensive evaluations to demonstrate their performance. Results show that \codename and \codenameplusplus outperform state-of-the-art fuzzers with more code coverage and more vulnerabilities discovered. The source code is available at: \url{https://github.com/zzyyrr/AlphaFuzz.git} and \url{https://github.com/zzyyrr/AlphaFuzzplusplus.git}.
\end{itemize}

\section{Motivation and insight}
In this section, we first illustrate our motivation by discussing the limitations of existing seed scheduling strategies with an example and then state our insight.

\subsection{Motivating example}\label{sec:motivation example}

\begin{figure}[ht]
  \centering
  \small
  \includegraphics[width=0.9\columnwidth]{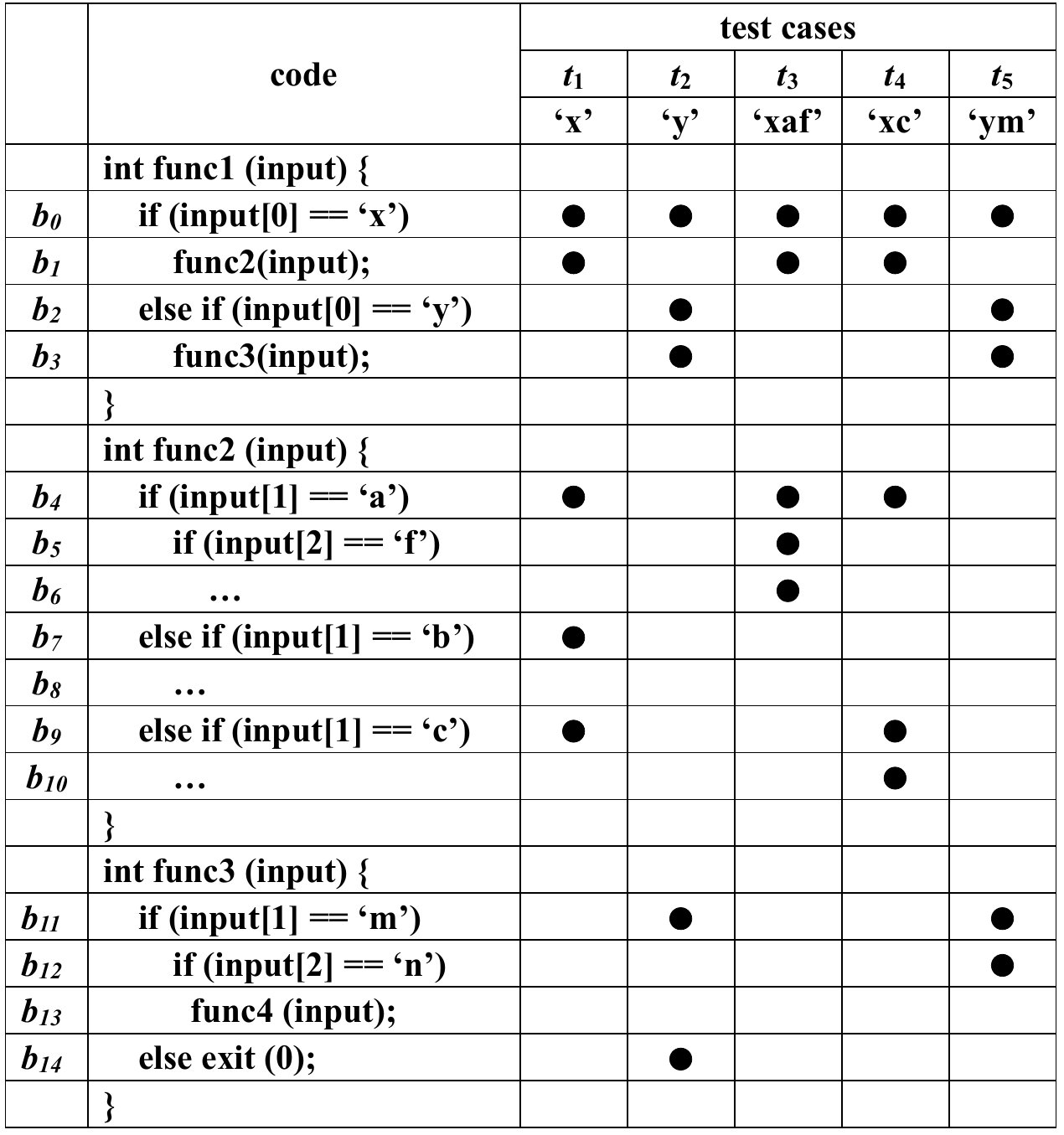}
  \caption{Motivating example.}
  \label{motivation}
\end{figure}

\begin{figure*}
  \centering
  \small
  \includegraphics[width=\textwidth]{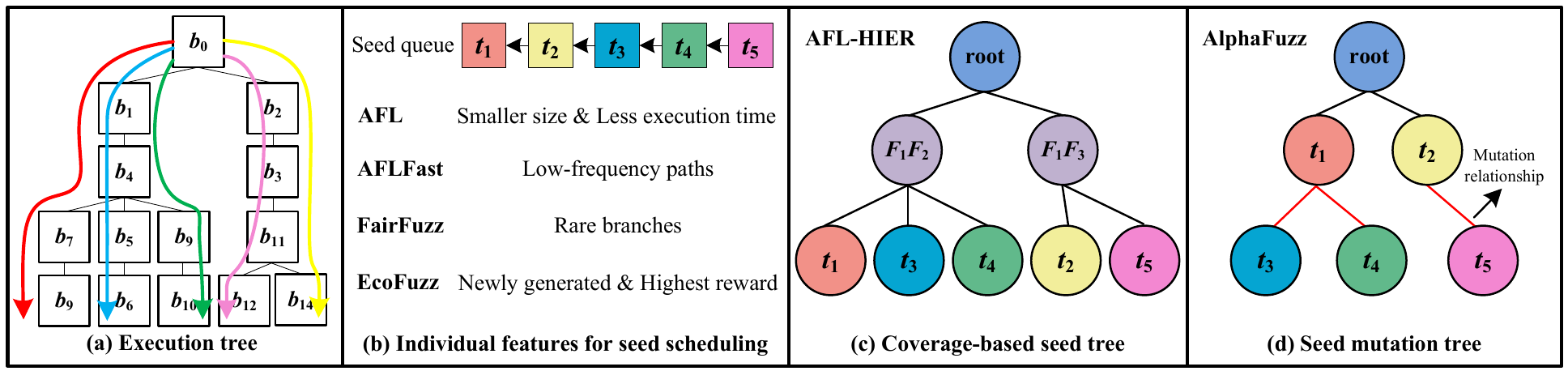}
  \caption{Existing seed scheduling strategies.}
  \label{seed-schedulers}
\end{figure*}

We use a piece of code in Figure~\ref{motivation} as an example to illustrate our motivation. 
As shown in Figure~\ref{motivation}, \emph{$b_i$} represents the basic block in this code, and $\left \langle b_i,\, b_j \right \rangle$ represents the branch from \emph{$b_i$} to \emph{$b_j$}.
In column 3 to 7, \emph{$t_1$} to \emph{$t_5$} are test cases retained by CGF. The string below each test case is the content of the input file. For example, the content of test case \emph{$t_1$} is $'x'$. When executing with different test cases, different basic blocks are covered. The solid circle indicates that the test case covers the corresponding basic block.

In order to describe the seed scheduling process clearly, we suppose that the CGF starts with two initial test cases, \emph{$t_1$} and \emph{$t_2$}.
After several fuzzing iterations, CGF generates and retains three new test cases. Test case \emph{$t_3$} and \emph{$t_4$} are generated via performing mutations on test case \emph{$t_1$}. Similarly, test case \emph{$t_5$} is derived from \emph{$t_2$}.

With these test cases in the seed pool, lots of seed scheduling strategies are proposed to guide the direction of CGF.
Figure~\ref{seed-schedulers} shows how existing seed scheduling strategies manage the test cases in Figure~\ref{motivation}. 
To better understand their principles, Figure~\ref{seed-schedulers} (a) shows the execution tree according to Figure~\ref{motivation}. 
Then we distinguish the paths covered by test cases with different colors.
For example, the path in blue color is covered by test case \emph{$t_3$}.

As shown in Figure~\ref{seed-schedulers} (b), \AFL~\cite{AFL}, \AFLFast~\cite{aflfast}, \FairFuzz~\cite{fairfuzz}, and \EcoFuzz~\cite{ecofuzz} propose different criteria for seed scheduling, but they all treat the seeds independently and only consider the individual features of seeds.
\AFL reserves the seed with the smallest size and shortest execution time for every covered branch to test as many inputs as possible in a limited time.
Since the seed \emph{$t_3$} covers the new branch $\left \langle  b_4,\, b_5\right \rangle$ and $\left \langle b_5,\, b_6\right \rangle$, and \emph{$t_4$} covers the new branch  $\left \langle  b_4,\, b_9\right \rangle$ and $\left \langle b_9,\, b_{10}\right \rangle$, they are both considered interesting, even though they both cover branch $\left \langle  b_0,\, b_1\right \rangle$ and $\left \langle  b_1,\, b_4\right \rangle$.
\AFLFast prefers low-frequency paths which fuzzing spent less time and effort exploring them than high-frequency paths.
Thus, \AFLFast chooses the paths covered by newly generated seeds \emph{$t_3$}, \emph{$t_4$}, and \emph{$t_5$}.
\FairFuzz calculates the times each branch has been covered and selects the seed that covers more low-frequency branches.
So, seeds \emph{$t_1$}, \emph{$t_3$}, and \emph{$t_4$} are selected by \FairFuzz.
\EcoFuzz gives the highest priority to unfuzzed seeds.
So \EcoFuzz selects seeds \emph{$t_3$}, \emph{$t_4$}, and \emph{$t_5$}.

The seed scheduling strategies above are unaware of the relationships among seeds.
Seeds covering new branches are always given high priority by the above seed selection strategies, regardless of weather the seeds are concentrated in the same code region.
For example, seeds \emph{$t_3$} and \emph{$t_4$} lead fuzzing to constantly explore the region of the \emph{func2}.
As a result, as long as new branches are consistently covered in this area of code, fuzzing will focus on this code region, and be difficult to jump out and explore other areas that may have potential vulnerabilities.

AFL-HIER~\cite{aflhier} defines many coverage metrics of different sensitivities to cluster the similar seeds together. 
As shown in Figure~\ref{seed-schedulers} (c), AFL-HIER uses function coverage and edge coverage to cluster the seeds together.
For example, seeds \emph{$t_1$}, \emph{$t_3$}, and \emph{$t_4$} are under the same node \emph{$F_1$,$F_2$} because they cover the same functions.
When performing seed scheduling, AFL-HIER starts from the nodes at function coverage level, and calculates scores for nodes based on their function-level rareness and their rewards.
The rewards of the cluster are the average rewards of all the seeds in this cluster.
However, elder seeds and newly generated seeds are clustered under the same nodes.
The rewards of newly generated seeds can be weakened by the fuzzing effort spent on elder seeds.
Therefore, the individual feature of seeds is flattened, which brings a negative impact on fuzzing deeper paths.

As discussed above, the seed scheduling strategy should consider both the relationships among seeds and the individual features of seeds. Therefore, we propose a new seed scheduling strategies which satisfy these two requirements.

\subsection{Our insight}
In this paper, we consider the seed scheduling as selecting an optimal path in order to explore its neighbor paths.
As discussed above, different paths in the execution tree lead fuzzing to explore different directions.
In addition, as the execution tree grows, some paths tend to focus on the same region of the program, while some paths tend to explore different regions.
The paths covered by seeds with mutation relationships share the same direction on the execution tree in the beginning, and then separate in different directions.
Therefore, we can leverage the mutation relationships among seeds to construct a ``seed mutation tree'', which can be an approximation of the execution tree.
As shown in Figure~\ref{seed-schedulers} (d), the relationships among different fuzzing directions are reserved in our ``seed mutation tree''.
For example, \emph{$t_1$}, \emph{$t_3$}, and \emph{$t_4$} are in the same sub-tree, because they all covered basic block \emph{$b_0$}, \emph{$b_1$} and \emph{$b_4$}. Similarly, \emph{$t_2$}, \emph{$t_5$} are under the same sub-tree.

In the meantime, the growth of the execution tree can also be represented in ``seed mutation tree'', where newly generated seeds are always leaf nodes and elder seeds are usually internal nodes.
Thus, the hierarchy of nodes in the tree can show the effort fuzzing spent on different paths, which is an important individual feature in seed scheduling, but is flattened in AFL-HIER. 

With this ``seed mutation tree'', seed scheduling can guide fuzzing process to exploit in a specific direction or explore in different directions.
To balance the exploration and exploitation, we model the seed scheduling as a Monte Carlo Tree Search (MCTS) problem~\cite{kocsis2006improved, browne2012survey, LiuWWGW20, BaierK20, Aleksander}. The following sections give the detail of our ``seed mutation tree'' and  MCTS-based seed scheduling strategy.

\section{Seed Mutation Tree}

In this section, we present the definition and the construction of the ``seed mutation tree''.

\subsection{Definition}
\label{sec:defination-of-tree}
We leverage the mutation relationships among seeds to construct the ``seed mutation tree'' as an approximation of the execution tree. 
The ``seed mutation tree'' is defined as follows.

\myparagraph{\bf Definition 1}. A {``seed mutation tree''} is a directed tree {\it T} = ({\it V}, {\it E}, {\it $\alpha$}), where:

$\bullet$ Each element {\it v} in the set of vertices {\it V} corresponds to a seed;

$\bullet$ Each element {\it e} in the set of edges {\it E $\subseteq$ V $\times$ V} corresponds to the {\em mutation relationship} between two vertices ${\it v}$ and ${\it w}$.

$\bullet$ The labeling function {\it $\alpha: E \rightarrow \Sigma$} associates edges among seeds in terms of their mutation relationships.

\subsection{Tree construction}
According to \textbf{Definition 1}, we construct a ``seed mutation tree'' during the fuzzing process.

First, before the fuzzing iteration begins, we construct the initial tree structure with the initial seeds and an auxiliary root node.
The auxiliary root node is the parent node of all the initial seeds for there might be more than one initial seed.
As shown in Figure~\ref{seed-schedulers} (d), the root node, node \emph{$t_1$} and node \emph{$t_2$} form the initial tree structure.

Then, CGF selects a seed as the base of mutation according to the seed scheduling strategy, and performs mutations on it to generate new test cases. 
We add the newly generated test cases that cover new paths of the program into the ``seed mutation tree'' as the child of the original seed. For example, CGF performs random mutations on seed \emph{$t_1$}, and generates test case \emph{$t_3$} and \emph{$t_4$}. They cover two new paths, \emph{$t_3$} covers the blue path and \emph{$t_4$} covers the green path, respectively.
We add the node \emph{$t_3$} to the tree as the child node of \emph{$t_1$} based on the mutation relationship between them, and node \emph{$t_4$} the same. 

As CGF continues to explore the targeted application, we continue to add the seeds covering new paths as nodes to the tree structure according to the mutation relationship between the seeds.

Some special mutations may operate on two seeds. For example, the mutation of splice in AFL will splice a seed with a second seed to generate a new input.
In such cases, we only construct an edge between the new node and the node corresponding to the first seed, because the first seed is selected by the seed scheduling strategy, whereas the second one is randomly selected for mutations.

\section{MCTS-based seed scheduling}
With ``seed mutation tree'', we regard the seed scheduling as searching for a seed along a tree, and propose an MCTS-based seed scheduling strategy. 
In this section, we first introduce the MCTS algorithm, and then present the design details of our MCTS-based seed scheduling strategy.

\subsection{Monte carlo tree search}
Monte Carlo Tree Search (MCTS)~\cite{kocsis2006improved, browne2012survey, LiuWWGW20, BaierK20, Aleksander} is a heuristic search algorithm for types of decision processes.
In principle, MCTS focuses on the most promising moves via expanding a search tree based on a random sampling of the search space. The process consists of four steps:

\myparagraph{Selection}. Using a specific selection strategy, usually UCT algorithm, MCTS starts from the root node $R$, recursively selects optimal child nodes until a leaf node $L$ is reached.

\myparagraph{Expansion}. If $L$ is not a terminal node, then MCTS creates one or more child nodes. Further, it will select one node $C$ from these child nodes. In this step, child nodes refer to any valid moves from the state defined by $L$.

\myparagraph{Simulation}. A simulation is performed by randomly choosing moves until a result or a predefined state is achieved.

\myparagraph{Back propagation}. This step back propagates from the new node $C$ to the root node $R$, and updates the simulation results.

\myparagraph{Upper confidence bounds for trees}.
The main challenge in a planning problem is to balance exploitation and exploration. For this challenge, the Upper Confidence Bounds (UCB) algorithm~\cite{ucb} is typically adopted.

\begin{equation}
    UCB = v_i + k\sqrt{\frac{lnN}{n_i}}
\label{ucb}
\end{equation}

The formula of UCB is shown in \textbf{Formula~\ref{ucb}}, where $v_i$ is the estimated value of the node, ${n_i}$ is the number of the times the node has been visited, $N$ is the total number of times that its parent has been visited, and $k$ is a tunable bias parameter.

The UCB formula balances the exploitation of known rewards with the exploration of relatively unvisited nodes for optimal performance. When applied in MCTS, the UCB formula is extended to the tree search and named as the Upper Confidence Bounds for Trees (UCT)~\cite{kocsis2006improved}. That is, UCT is a special case for UCB in MCTS.

\subsection{Tree search for seed scheduling}
Inspired by the MCTS algorithm, our MCTS-based seed scheduling strategy can calculate and set a score for every node, with both fuzzing rewards as well as fuzzing effort for balancing exploration and exploitation. 
However, we cannot directly employ MCTS for searching a seed, because the MCTS algorithm always selects the leaf node.
In our ``seed mutation tree'', an internal node can also be selected as a seed even if it has multiple child nodes. With this impact, the MCTS algorithm is incompatible with ``seed mutation tree''.

\begin{figure}
  \centering
  \small
  \includegraphics[width=0.5\columnwidth]{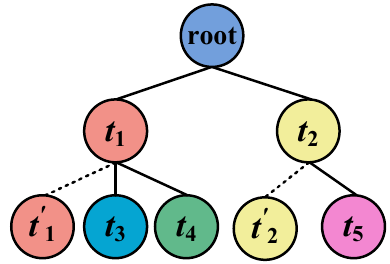}
  \vspace{-1.0em}
  \caption{``Seed mutation tree'' with variants.}
  \label{variants-tree}
  \vspace{-0.2in}
\end{figure}

\begin{figure*}
  \centering
  \small
  \includegraphics[width=\textwidth]{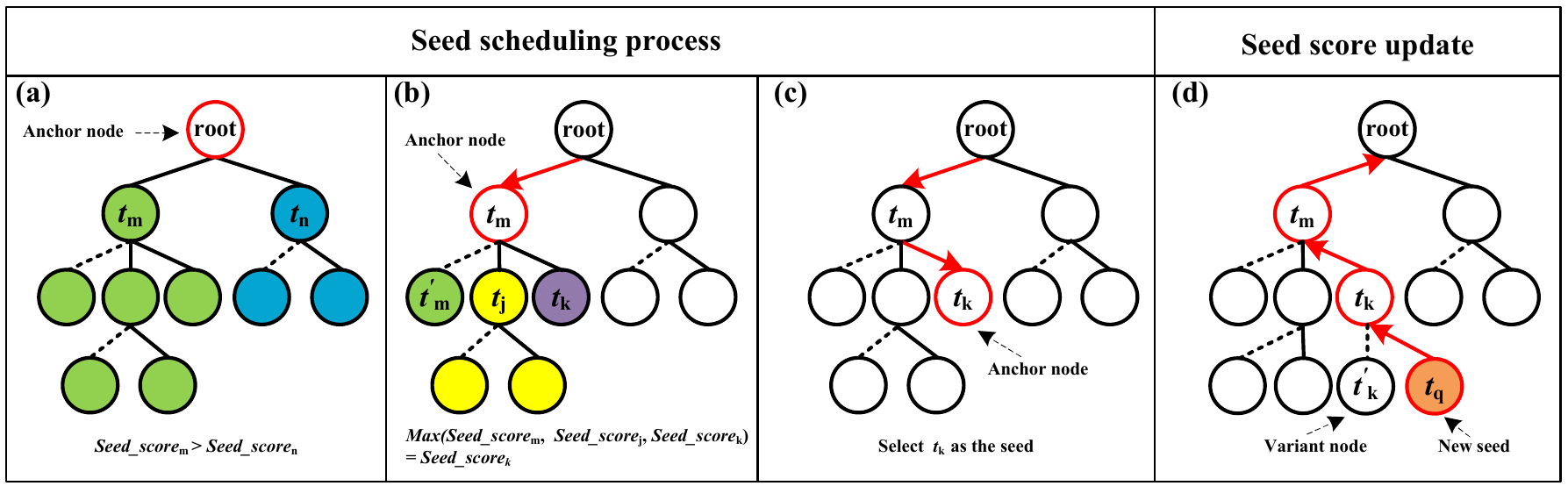}
  \vspace{-2em}
  \caption{MCTS-based seed scheduling process.}
  \vspace{-1.5em}
  \label{seed-scheduling-steps}
\end{figure*}

\subsubsection{\textbf{Variants for internal nodes}}
To address the problem of incompatible, we update the ``seed mutation tree'' by inserting a variant node for every internal node.
In terms of seed scheduling, there are two different roles for an internal node. First, an internal node refers to a seed. Second, an internal node is also the root node of a sub-tree, from the perspective of the tree structure. 

More important, these two different roles indicate different fuzzing scores. As a seed, it refers to the fuzzing score for selecting this seed. As the root node of a sub-tree, it refers to the summary fuzzing score for selecting every seed in the sub-tree.
With the impact of such two different roles, the calculations of fuzzing score for an internal node will be also different.

For this problem, we add a variant for every internal node. The internal node refers to the root node of its sub-tree, and the variant refers to the corresponding seed.
In this way, the variant will be a leaf node of the internal node. 
Regarding the fuzzing scores, the internal node delivers the summary fuzzing scores for every seed in the sub-tree. By contrast, the variant delivers the fuzzing score for the corresponding seed.

Let us take the motivating example for illustration. 
As show in Figure~\ref{variants-tree}, $t_1$ has two child nodes, $t_3$ and $t_4$. $t_2$ has one child node $t_5$. To distinguish between an internal node itself and its sub-tree, we construct two variants for both $t_1$ and $t_2$, which are denoted as $t_1'$ and $t_2'$, respectively.

With variants, the construction of ``seed mutation tree'' requires updating for internal nodes. Whenever a leaf node is generated from a node without a variant node, a variant of the parent node will also be created. Then, both the leaf node and the variant node will be inserted into the tree as the child nodes of the parent node. 

The design of variants addresses the problem of incompatibility between MCTS and ``seed mutation tree''.
By setting a variant for every internal node, the variant node will be a leaf node for the internal node.
With this design, MCTS is able to search for a promising node until a leaf node is reached.

\subsubsection{\textbf{Seed-mutation-tree search}}
The process of our seed scheduling strategy is shown in Figure~\ref{seed-scheduling-steps}.
Each fuzzing iteration starts from the root node, we label the currently selected node as the anchor node. 
Then, we calculate $seed\_scores$ for every child node of the anchor node. 
We will present the calculation formula in \cref{sec:UCT-algorithm}.
For example, we calculate the $seed\_scores$ for \emph{$t_m$} and \emph{$t_n$}.
If the $seed\_scores$ of \emph{$t_m$} is higher than \emph{$t_n$}, the anchor node moves to \emph{$t_m$}.

As shown in Figure~\ref{seed-scheduling-steps} (b), by comparing $seed\_scores$ between sibling nodes, our MCTS-based seed scheduling strategy will recursively select the promising one with the highest score, until a leaf node is reached. 
As shown in Figure~\ref{seed-scheduling-steps} (c), \emph{$t_k$} is a leaf node, therefore the seed scheduling stops and selects \emph{$t_k$} as the base for the mutation stage of fuzzing.

\subsection{Seed quantification using UCT algorithm}\label{sec:UCT-algorithm}
We leverage the UCT algorithm to calculate the score of a seed in terms of the effort and the rewards.
We evaluate the fuzzing effort that cost by fuzzing the same seed rapidly and calculate the rewards of a scheduled seed in terms of new code coverage traversed by the mutations based on this seed. With these two factors, fuzzing effort and fuzzing rewards, we design the calculation of $seed\_score$ as \textbf{Formula~\ref{F_seed-score}}.

\begin{equation}
    seed\_score = \frac{q_i}{n_i}+k\sqrt{\frac{lnN_i}{n_i}}
\label{F_seed-score}
\end{equation}

In \textbf{Formula~\ref{F_seed-score}}, $n_i$ records the number of times this node has been selected.  $q_i$ represents the size of the difference set between the branches covered by the current seed and the branches covered by other seeds under the same sub-tree. We will describe the calculation of $q_i$ in detail later.
In $k\sqrt{\frac{lnN_i}{n_i}}$, $k$ is a constant, we discuss the value of $k$ in \textbf{\cref{sec:parameter_k}}. $N_i$ represents the number of times the parent of the node has been scheduled.

We can see that, $\frac{q_i}{n_i}$ represents the mean rewards of this seed.
The $k\sqrt{\frac{lnN_i}{n_i}}$ represents the fuzzing effort spent on this seed, we can observe that the increasing value of $n_i$ will decrease $seed\_score$, while the increasing of $q_i$ will increase $seed\_score$.

We calculate $q_i$ for every child node of the anchor node. 
As shown in Figure~\ref{seed-scheduling-steps} $(b)$, \emph{$t_m$} is the anchor node, the child nodes of \emph{$t_m$} have three types, variant node \emph{$t'_m$}, internal node \emph{$t_j$} and leaf node \emph{$t_k$}.
The variant node represents the path covered by the anchor node. 
The leaf node \emph{$t_k$} represents the path covered by seed \emph{$t_k$}.
The internal node \emph{$t_j$} represents a group of paths covered by all the nodes in the sub-tree which are the yellow nodes.

Most of the existing fuzzers leverage branch coverage to guide the evolution.
We choose a node which covers more \emph{unique branches}.
Specifically, for each variant node and leaf node, we collect the branches covered by the seed as its \emph{branch set}, respectively.
For the internal node, we count the branches covered by this node and all its descendants as its \emph{branch set}.
We then put these \emph{branch sets} together and calculate the difference set for each child node.
The number of the branches in the difference set means the number of \emph{unique branches} covered by the child node compared to other child nodes.
Then, we assign the number of \emph{unique branches} to the value of $q_i$.

\subsection{Seed score update}
After each round of seed scheduling, we update the rewards and effort from the selected seed back to the root node.

To update the fuzzing rewards, as shown in the Figure~\ref{seed-scheduling-steps} (d) , when a new node, \emph{$t_q$} for example, is added to a sub-tree, we update the information of newly covered branches of \emph{$t_q$} from \emph{$t_q$} to all its parent nodes \emph{$t_i$} and \emph{$t_m$}.

To update the fuzzing effort, the selected times of the selected node and all its parent nodes are increased by one. That is to say, the selected times of \emph{$t_i$} and \emph{$t_m$} are increased by 1.
This process goes in exactly the opposite direction to seed scheduling.
In this way, we are able to record the rewards and effort of this time's seed scheduling to guide the next iteration.

\section{Evaluation}
We implement two prototypes, \codename and \codenameplusplus, on top of AFL~\cite{AFL} and AFL++~\cite{AFLplusplus-Woot20}, respectively. 
To demonstrate the effectiveness of our approach, we design and conduct experiments to answer the following research questions:
\begin{itemize}
\item \noindent \textbf{RQ1.} How does our approach perform in vulnerability detection on benchmark datasets?

\item \noindent\textbf{RQ2.} How significant is the improvement of code coverage compared to baseline fuzzing techniques?

\item \noindent\textbf{RQ3.} How does our approach affect fuzzing throughput?

\item \noindent\textbf{RQ4.} How does the parameter $k$ in \textbf{Formula~\ref{F_seed-score}} affect fuzzing performance?

\item \noindent\textbf{RQ5.} Can \codename detect new vulnerabilities in real-world applications?

\end{itemize}

\subsection{Datasets}
We leverage three datasets: the Cyber Grand Challenge (CGC) dataset~\cite{DARPA}, UniFuzz~\cite{UniFuzz}, and 12 real-world binaries.

The CGC dataset~\cite{DARPA} includes binaries from the CGC Qualifying Event and the CGC Final Event, which are widely used in previous techniques~\cite{aflhier,redqueen}.
Every CGC binary is injected with one or more memory corruption vulnerabilities.
In our evaluation, we exclude programs involving communication with multiple programs, and programs on which AFL cannot work. In total, we use 188 CGC binaries for evaluation. As the original CGC binaries are customized with specific syscalls, we leverage their compatible Linux versions~\cite{multios} for evaluation.

UniFuzz~\cite{UniFuzz} is an open-source and pragmatic metrics-driven platform for evaluating fuzzers, which consists of 20 real-world programs. Unfortunately, \emph{sqlite3} and \emph{ffmpeg} failed to run correctly due to frequent timeouts during our evaluation. Thus, we only leverage 18 programs.

\begin{table}
\centering
\setlength{\belowcaptionskip}{1.5mm}
\caption{Real-world binaries evaluated in our experiments.}
\resizebox{\columnwidth}{!}{
\begin{tabular}{l l l l l l}
    \toprule
    \textbf{Program} & \textbf{Input} & \textbf{cmd line} & \textbf{Program} & \textbf{Input} & \textbf{cmd line} \\
    \midrule
    cjpeg & bmp &  @@ & infotocap & text & @@ \\
    exiv2 & jpg &  @@ /dev/null &  mp3gain & mp3 &  @@ \\
    nm & elf & -AD @@ & objdump & elf & -d @@ \\
    pdfimages & pdf & @@ /dev/null & pngfix & png & @@ \\
    readelf & elf & -a @@ & size & elf & -At @@ \\
    tiff2pdf & tiff & @@ & xmlwf & xml & @@\\
    \bottomrule
\end{tabular}
}
\label{Table-linux}
\end{table}

To answer the \textbf{RQ5}, we choose 12 real-world binaries based on the following features: popularity, frequency of being tested, and diversity of categories. As shown in Table~\ref{Table-linux}, these 12 real-world binaries include popular tools (e.g., \emph{nm}, \emph{objdump}), image processing libraries (e.g., \emph{libjpeg}, \emph{libtiff}), terminal processing libraries (e.g., \emph{ncurses}), and document processing libraries (e.g., \emph{xpdf}), etc.

\subsection{Baseline techniques}
In recent years, a number of fuzzing techniques have been proposed to improve the performance of fuzzing from different aspects, including coverage sensitivity, mutation algorithms, input generation, and execution monitoring~\cite{collafl, mopt, Angora, SAFL, skyfire, perfuzz, untracer, PTFUZZ}. As \codename proposes a new seed scheduling strategy, we only select fuzzing techniques focusing on seed scheduling algorithms as baseline techniques. Besides, we exclude fuzzing techniques requiring the source code of the targeted binaries. 
For AFL-based fuzzing techniques, we choose \AFL~\cite{AFL}, \AFLFast~\cite{aflfast}, \FairFuzz~\cite{fairfuzz} and \EcoFuzz~\cite{ecofuzz}. As AFL-HIER~\cite{aflhier} only releases their prototype as AFL++-HIRE, which is implemented on top of AFL++~\cite{AFLplusplus-Woot20}, thus we further choose AFL++ and AFL++-HIRE as baseline techniques.

\subsection{Experiment setup}
We run the experiments on a server configured with 40 CPU cores of 2.50GHz E5-2670 v2, 125GB RAM, and running on the 64-bit Ubuntu 16.04 LTS.

We fuzz each binary for 24 hours as done in previous studies~\cite{aflfast,fairfuzz}. To alleviate the impact of the nature of randomness in fuzzing, we run each experiment for 10 rounds, and report the statistical results for a more comprehensive evaluation.

All the experiments are based on the QEMU-mode in AFL (so as to support binary-only targets) and configured with the same initial seeds and instructions. The initial seeds for CGC binaries come from the examples provided by the CGC challenges~\cite{DARPA}. The initial seeds for UniFuzz are provided by UniFuzz~\cite{UniFuzz}. The initial seeds for the 12 real-world binaries come from the default seed examples provided by \AFL. 

\subsection{Vulnerability detection}
To measure the vulnerability detection capability of \codename, we analyse the experimental results on CGC and UniFuzz. As CGC binaries are manually designed and embedded with known bugs, we mainly evaluate the number of unique bugs and the time to first crash. For UniFuzz, we evaluate the number of unique bugs and the number of unique exploitable vulnerabilities.

\subsubsection{\textbf{Unique bugs on CGC dataset}}
To summarize the evaluation results of the 10-round experiments, we count the number of times each bug is found in 10-round experiments.
Then we count the number of bugs in terms of the number of times the bug is found in 10-round experiments by each baseline technique.
As shown in Table~\ref{Table-cgc-crashes}, columns 2-11 list the number of bugs that are discovered in 10 rounds by different metrics. Column 2 reports the number of vulnerabilities that are discovered in every round. Column 3 reports the number of vulnerabilities that are discovered in at least 9 rounds, and so on. 

\begin{table}
    \centering
    \setlength{\belowcaptionskip}{1.5mm}
	\caption{Numbers of discovered bugs on CGC.}
    \label{Table-cgc-crashes}
    \resizebox{\columnwidth}{!}{
    		\begin{tabular}{ l | c | c | c | c | c | c | c | c | c | c }
    	\toprule
		\bf{Fuzzer}  & $\bf{= 10}$ & $\bf{\ge{9}}$ & $\bf{\ge{8}}$ & $\bf{\ge{7}}$ & $\bf{\ge{6}}$ & $\bf{\ge{5}}$ & $\bf{\ge{4}}$ & $\bf{\ge{3}}$ & $\bf{\ge{2}}$ & $\bf{\ge{1}}$ \\
		  \midrule
			\codename & 70 & 72 & 72 & 73 & 75 & 75 & 80 & 83 & 87 & 87 \\
			\AFL & 69 & 69 & 69 & 70 & 71 & 75 & 78 & 83 & 83 & 83 \\
			\AFLFast & 67 & 67 & 70 & 71 & 73 & 77 & 79 & 79 & 81 & 83 \\
			\FairFuzz & 67 & 67 & 68 & 70 & 71 & 72 & 72 & 73 & 74 & 76 \\
			\EcoFuzz & 66 & 66 & 66 & 67 & 68 & 68 & 68 & 68 & 73 & 73 \\
            \midrule
            \codenameplusplus & 72 & 73 & 75 & 75 & 77 & 80 & 80 & 85 & 88 & 88 \\
            
            AFL++ & 70 & 72 & 75 & 75 & 75 & 76 & 79 & 83 & 83 & 84 \\
           
            AFL++\-Hier & 72 & 73 & 73 & 75 & 76 & 80 & 83 & 83 & 85 & 85 \\
        \bottomrule
		\end{tabular}
		}
\end{table}

From Table~\ref{Table-cgc-crashes}, we can observe that in nearly all these 10 metrics, \codename discovers more bugs than the other AFL-based fuzzing techniques.
For example, \codename discovers 87 bugs through 10 rounds, with 4 more bugs than \AFL and \AFLFast, 11 more bugs than \FairFuzz, and 14 more bugs than \EcoFuzz.
By examining the numbers of discovered bugs in Table~\ref{Table-cgc-crashes}, we observe that the performance of \AFL is almost the same as \AFLFast on CGC binaries.
On the contrary, \FairFuzz and \EcoFuzz work worse than other fuzzing techniques on CGC binaries.

For AFL++-based fuzzing techniques, in nearly all these 10 metrics, \codenameplusplus discovers more bugs than AFL++ and AFL++-HIER. \codenameplusplus discovers 88 bugs through 10 rounds, with 4 more bugs than AFL++ and 3 more bugs than AFL++-HIER.

\begin{figure}
\centering
\includegraphics[width=0.9\columnwidth]{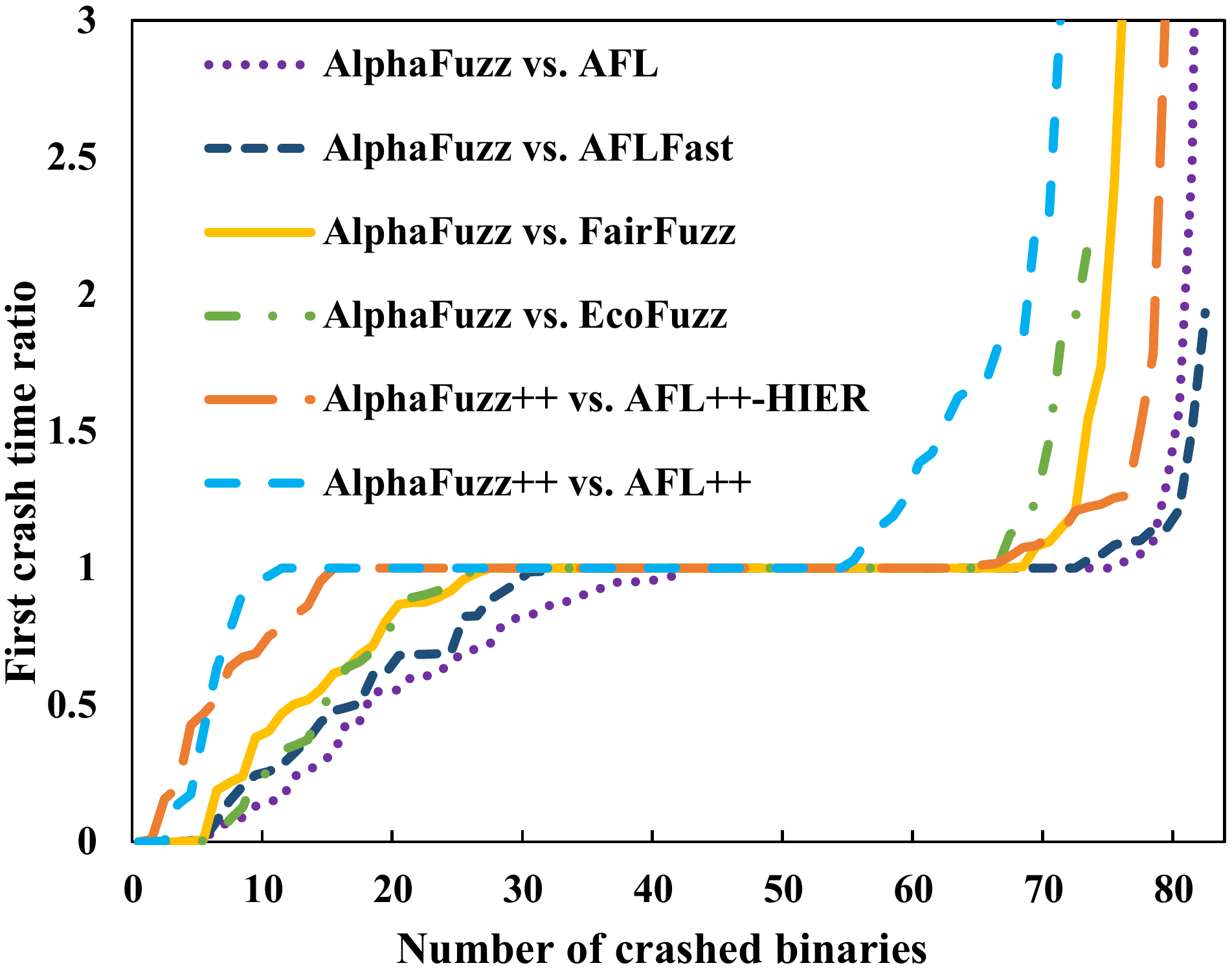}
\vspace{-1em}
\caption{First time to crash.}
\label{First time to crash}
 \vspace{-1.5em}
\end{figure}

\subsubsection{\textbf{The time to first crash on CGC dataset.}}
For CGC binaries, we further leverage the time of first crash to evaluate the performances of fuzzing techniques as the previous technique AFL-HIER~\cite{aflhier}.
The time of first crash refers to the time costed by a fuzzing technique to detect crashes. A fuzzing technique that discovers crashes more quickly than another technique is said to perform better.
To calculate the time to first crash, we first calculate the average time it takes for a binary to crash during our 10-round experiments.
To ease representation, we compare the performance for each pair of fuzzing techniques in terms of the time to first crash. For example, taking AFL as the baseline, we compare \codename with AFL by dividing the time to first crash in \codename by the time to first crash in AFL. If the ratio is less than 1, it means that \codename detects crashes faster than AFL.

Figure~\ref{First time to crash} shows the first crash time comparison for each pair of fuzzing techniques. Take the \codenameplusplus vs. AFL++ for illustration (the light blue line in Figure~\ref{First time to crash}). Both \codenameplusplus and AFL++ detect crashes in 82 binaries. Among these crashes, the first crash time ratio for 55 of them is no larger than 1, which indicates that \codenameplusplus spend less effort or the same on 55 out of the 82 binaries to discovery the vulnerability than AFL++. 

From Figure~\ref{First time to crash}, we can observe that \codename finds crashes faster than other baseline techniques. In detail, \codename finds crashes faster than \AFL on 50.6\% of programs. \codename finds crashes faster or as fast as \AFL on 91.6\% of programs.
Similarly, we can observe that \codenameplusplus detects more crashes than AFL++ and AFL++-HIER within given fuzzing time. 

\begin{table}
\renewcommand\arraystretch{0.7}
    \setlength{\belowcaptionskip}{1.5mm}
    \caption{Numbers of discovered unique bugs.}
    \label{Table-bug-UniFuzz}
   \resizebox{\columnwidth}{!}{
    \begin{tabular}{l|c|c|c|c|c}
    \toprule
        \textbf{Binary} & \textbf{\AFL} & \textbf{\AFLFast} &  \textbf{\EcoFuzz} & \textbf{\FairFuzz} & \textbf{\codename} \\
    \midrule
        cflow &	3 &	4 &	2 &	3 &	4\\
        exiv2 &	0 &	1 &	1 &	1 &	1 \\
        flvmeta &	3 &	3 &	3 &	4 &	3 \\
        gdk &	6 &	6 &	2 &	0 &	8 \\
        imginfo &	0 &	1 &	0 &	0 &	1 \\
        infotocap &	2 &	3 &	0 &	2 &	2 \\
        jhead &	0 &	0 &	2 &	0 &	0 \\
        jq &	0 &	0 &	0 &	0 &	0 \\
        lame &	3 &	2 &	2 &	3 &	3 \\
        mp3gain &	6 &	7 &	4 &	3 &	7 \\
        mp42aac &	2 &	2 &	2 &	2 &	2 \\
        mujs &	0 &	0 &	0 &	0 &	1 \\
        nm &	0 &	0 &	0 &	0 &	0 \\
        objdump &	3 &	3 &	2 &	1 &	4 \\
        pdftotext &	1 &	1 &	0 &	1 &	8 \\
        tcpdump &	0 &	0 &	0 &	0 &	0 \\
        tiffsplit &	4 &	5 &	3 &	8 &	6 \\
        wav2swf &	2 &	3 &	2 &	2 &	3 \\
    \midrule
    \bf{total} &	\bf{35} &	\bf{41} &	\bf{25} &	\bf{30} &	\bf{53} \\
    \bottomrule
    \end{tabular}
    }
\end{table}
\subsubsection{\textbf{Unique bugs on UniFuzz.}} For the UniFuzz dataset, we first calculate the number of unique bugs discovered by each fuzzing technique. According to UniFuzz~\cite{UniFuzz}, we leverage the report produced by ASAN~\cite{ASAN} and GDB~\cite{GDB} to de-duplicate bugs. 

Table~\ref{Table-bug-UniFuzz} shows the numbers of detected unique bugs by AFL-based fuzzing techniques. Among the 18 binaries, 3 of them have not been detected bugs by any fuzzing techniques. We can observe that \codename finds no less unique bugs than other fuzzing techniques on 11 out of the 15 binaries. In total, \codename finds 18, 12, 28, and 23 more unique bugs than \AFL, \AFLFast, \EcoFuzz, and \FairFuzz, respectively.

Table~\ref{Unique-exploitable-alphuzz++} shows the numbers of detected unique bugs by AFL++-based fuzzing techniques. Among the 18 binaries, 2 of them have not been detected bugs by any fuzzing techniques. We can observe that \codenameplusplus finds no less unique bugs than AFL++ on 14 binaries, and no less unique bugs than AFL++-HIER on 12 binaries.

\begin{table}
\renewcommand\arraystretch{0.7}
    \setlength{\belowcaptionskip}{1.5mm}
    \caption{Number of discovered unique exploitable bugs.}
    \label{Table-exploitable-UniFuzz}
    \resizebox{\columnwidth}{!}{
    \begin{tabular}{l|c|c|c|c|c}
    \toprule
    \textbf{Binary} & \textbf{\AFL} & \textbf{\AFLFast} &  \textbf{\EcoFuzz} & \textbf{\FairFuzz} & \textbf{\codename} \\
    \midrule
        cflow &	0 &	0 &	0 &	0 &	0\\
        exiv2 &	0 &	1 &	1 &	1 &	1 \\
        flvmeta &	1 &	1 &	1 &	1 &	1 \\
        gdk &	4 &	3 &	2 &	0 &	3 \\
        imginfo &	0 &	0 &	0 &	0 &	0 \\
        infotocap &	0 &	0 &	0 &	0 &	0 \\
        jhead &	0 &	0 &	2 &	0 &	0 \\
        jq &	0 &	0 &	0 &	0 &	0 \\
        lame &	2 &	1 &	1 &	1 &	2 \\
        mp3gain &	1 &	1 &	0 &	0 &	1 \\
        mp42aac &	0 &	0 &	0 &	0 &	0 \\
        mujs &	0 &	0 &	0 &	0 &	0 \\
        nm &	0 &	0 &	0 &	0 &	0 \\
        objdump &	2 &	2 &	0 &	1 &	2 \\
        pdftotext &	0 &	1 &	0 &	0 &	3 \\
        tcpdump &	0 &	0 &	0 &	0 &	0 \\
        tiffsplit &	1 &	1 &	1 &	1 &	1 \\
        wav2swf &	2 &	3 &	2 &	2 &	3 \\
    \midrule
    \bf{total} & \bf{13} & \bf{14} & \bf{10} & \bf{6} &	\bf{17} \\
    \bottomrule
    \end{tabular}
    }
\end{table}

\begin{table}
\renewcommand\arraystretch{1}
    \caption{Number of discovered unique bug and exploitable vulnerabilities found by AFL++-based techniques.}
    \label{Unique-exploitable-alphuzz++}
    \resizebox{\columnwidth}{!}{
    \begin{tabular}{l|c|c|c|c|c|c}
    \toprule
    \multirow{2}{1cm}{\bf{Binary}} & \multicolumn{3}{c|}{\bf{Unique bugs}} & \multicolumn{3}{c}{\bf{Exploitable bugs}} \\
    \cline{2-7}{}  & \bf{AFL++} & \bf{AFL++-HIER} & \bf{\codenameplusplus} & \bf{AFL++} & \bf{AFL++-HIER} & \bf{\codenameplusplus}\\
    
    \midrule
        cflow & 3 & 5 & 5 & 1 & 1 & 1 \\
        exiv2 & 2 & 2 & 2 & 1 & 1 & 1\\
        flvmeta & 4 & 3 & 4  & 1 & 1 & 1\\
        gdk & 10 & 9 & 12 & 6 & 3 & 8\\
        imginfo & 0 & 1 & 1 & 0 & 0 & 0\\
        infotocap & 2 & 0 & 3 & 0 & 0 & 0\\
        jhead & 3 & 3 & 3 & 0 & 0 & 0\\
        jq & 0 & 0 & 0 & 0 & 0 & 0\\
        lame & 2 & 3 & 3 & 2 & 2 & 2\\
        mp3gain & 6 & 6 & 6 & 1 & 1 & 1 \\
        mp42aac & 2 & 2 & 2 & 0 & 0 & 0\\
        mujs & 0 & 0 & 1 & 0 & 0 & 0\\
        nm  & 0 & 0 & 0 & 0 & 0 & 0\\
        objdump & 2 & 3 & 3 & 2 & 1 & 2\\
        pdftotext & 2 & 8 & 4 & 2 & 2 & 3\\
        tcpdump & 0 & 0 & 1 & 0 & 0 & 0\\
        tiffsplit &  3 & 6 & 5 & 1 & 1 & 1\\
        wav2swf & 3 & 2 & 4 & 3 & 2 & 4\\
    \midrule
    \bf{total} & \bf{44} & \bf{55}& \bf{59}& \bf{20}& \bf{15}& \bf{24}\\
    \bottomrule
    \end{tabular}
    
}
    \vspace{-1.5em} 
\end{table}

\begin{table}
\small
        \begin{center}
        \setlength{\belowcaptionskip}{1.5mm}
        \caption{$p$ values of the code coverage on UniFuzz with \codename as the baseline.}
        \vspace{-1.5em}
        \label{pvalue}
        \resizebox{\columnwidth}{!}{
        \begin{tabular}{ l | c | c | c | c | c }
            \toprule
            \multirow{2}{1cm}{\bf{Binary}} & \multicolumn{1}{c|}{\bf{\codename}} & \multicolumn{1}{c|}{\bf{\AFL}} & \multicolumn{1}{c|}{\bf{\AFLFast}} & \multicolumn{1}{c|}{\bf{\EcoFuzz}} & \multicolumn{1}{c}{\bf{\FairFuzz}}\\
            \cline{2-6}  & \bf{AVG cov} & \bf{$p$ value} & \bf{$p$ value} & \bf{$p$ value} & \bf{$p$ value} \\
            \midrule
            cflow & 4.859\% & \emph{\textbf{0.0209}} & \emph{\textbf{0.0408}} & 0.4795 & \emph{\textbf{0.0078}}\\
            exiv2 & 15.728\% & \emph{\textbf{0.0047}} & \emph{\textbf{0.0023}} & 0.0511 & \emph{\textbf{0.0054}} \\
            flvmeta & 2.33 \% & \textgreater 0.10000 & \emph{\textbf{0.0001}} & 0.0867 & \textgreater 0.9999  \\
            gdk & 9.05\%  & \emph{\textbf{\textless 0.0001}} & \emph{\textbf{0.0014}} & \emph{\textbf{0.0014}} & \emph{\textbf{0.0014}} \\
            imginfo & 8.324\% & \emph{\textbf{0.0002}} & \emph{\textbf{0.0047}} & \emph{\textbf{0.0082}} & \emph{\textbf{0.0002}} \\
            infotocap & 5.267\% & 0.4221 & 0.3046 & \emph{\textbf{\textless 0.0001}} & \emph{\textbf{0.0027}} \\
            jhead & 1.477\% & \textgreater 0.9999 & 0.0698 & \emph{\textbf{0.0031}} & \emph{\textbf{0.0031}}\\
            jq & 8.34\% & \emph{\textbf{0.0435}} & 0.8951 & \emph{\textbf{0.022}} & 0.1898\\
            lame & 14.161\% & 0.4648 & 0.2387 & 0.1008 & 0.4692 \\
            mp3gain & 3.891\% & 0.9887 & 0.4027 & 0.1342 & 0.1013\\
            mp42aac & 7.734\% & \emph{\textbf{0.0002}} & \emph{\textbf{\textless 0.0001}} & \emph{\textbf{\textless0.0001}} & \emph{\textbf{\textless0.0001}} \\
            mujs & 12.984\% & 0.3047 & \emph{\textbf{\textless 0.0001}} & \emph{\textbf{\textless 0.0001}} & \emph{\textbf{\textless0.0001}} \\
            nm & 6.064\% & 0.1172 & \emph{\textbf{0.0209}} & \emph{\textbf{\textless 0.0001}} & \emph{\textbf{0.0217}}\\
            objdump & 11.654\% & 0.8978 & 0.3047 & 0.4699 & 0.4695 \\
            pdftotext & 25.904\% & 0.6415 & \emph{\textbf{0.0024}} & 0.3035 & 0.9126\\
            tcpdump & 22.204\% & \emph{\textbf{0.0009}} & \emph{\textbf{\textless 0.0001}} & \emph{\textbf{\textless 0.0001}} & \emph{\textbf{\textless0.0001}}\\
            tiffsplit & 5.231\% & 0.072 & 0.3227 & 0.7811 & \emph{\textbf{0.0002}} \\
            wav2swf & 1.39\% & 0.3067 & \emph{\textbf{0.0086}} & \textgreater 0.9999 & 0.3043\\
            \bottomrule
        \end{tabular}
        }
    \end{center}
\vspace{-1.5em}
\end{table}

\subsubsection{\textbf{Exploitable vulnerabilities on UniFuzz}}
Exploitability reflects the severity of the vulnerability~\cite{UniFuzz}. 
In this paper, we leverage the GDB Exploitable~\cite{exploitable} to identify exploitable vulnerabilities. GDB Exploitable classifies crashes into four categories: EXPLOITABLE, PROBABLY\_EXPLOITABLE, PROBABLY\_NOT\_EXPLOITABLE and UNKNOWN. Table ~\ref{Table-exploitable-UniFuzz} and Table~\ref{Unique-exploitable-alphuzz++} show the number of crashes that are classified as EXPLOITABLE and PROBABLY\_EXPLOITABLE. We can observe that \codename finds 4, 3, 7, and 11 more exploitable vulnerabilities than AFL, AFLFast, EcoFuzz, and FairFuzz, respectively. \codenameplusplus finds 4 and 9 more exploitable vulnerabilities than AFL++ and AFL++-HIER, respectively.

\subsection{Code coverage}
Code coverage is a critical metric for evaluating the performance of a fuzzing technique~\cite{Jinghan}. Basically, the more code a fuzzing technique can cover, the more likely it is to find the hidden bugs. We use the bitmap maintained by \AFL, a variant of branch coverage, to measure the code coverage. Specifically, \AFL maps each branch transition into an entry of the bitmap via hashing. If a branch transition is explored, the corresponding entry in the bitmap will be filled and the size of the bitmap will increase. 

\begin{figure*}
\centering
\includegraphics[width=\textwidth]{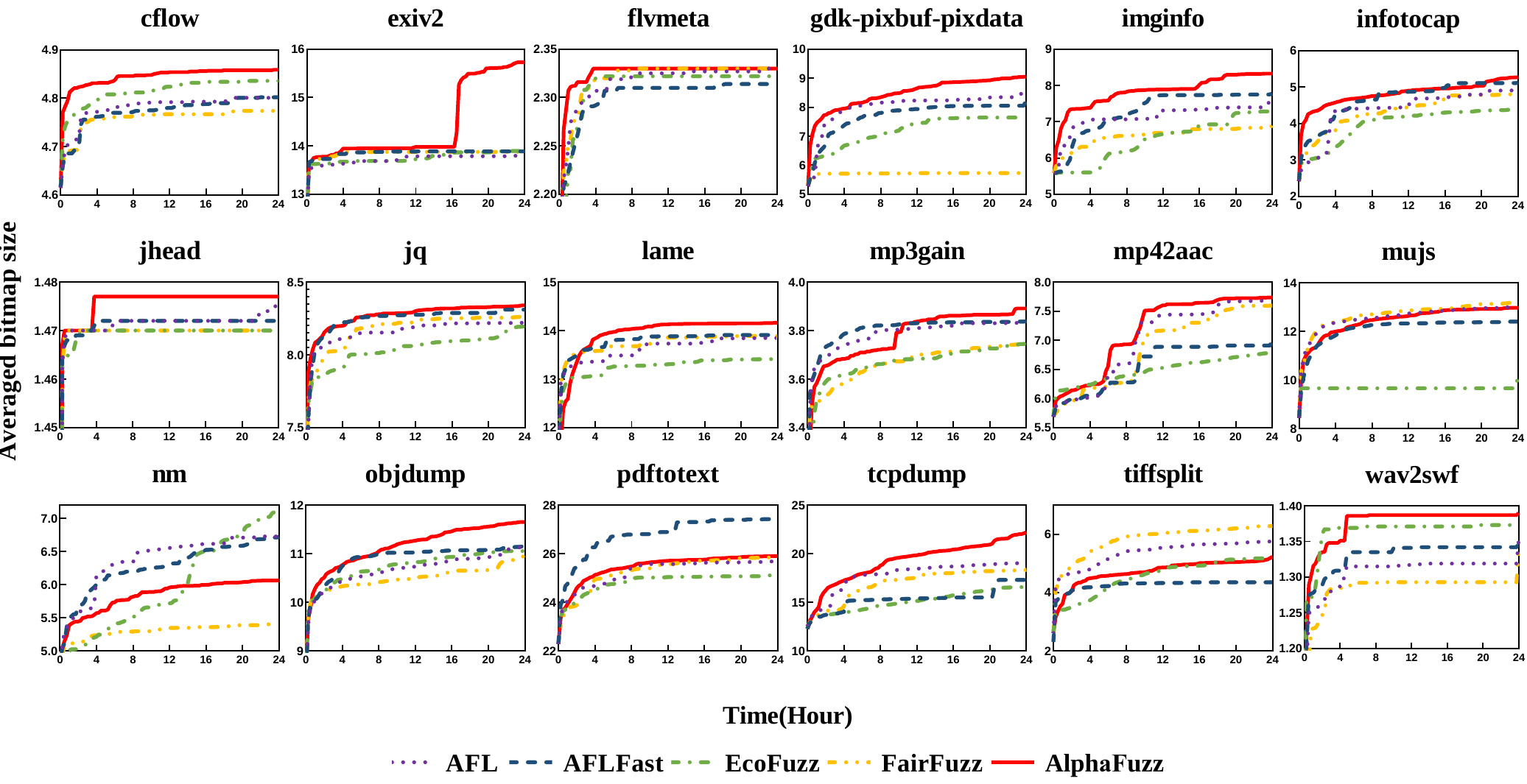}
\caption{Average bitmap size for every binary of UniFuzz. Each graph represents a binary, with the abscissa representing the time and the ordinate representing the size of the average bitmap size.}
\label{Figure-every-bitmap}
\end{figure*}

As an in-depth analysis, we present the average bitmap size of the 10-round experiments for every binary in UniFuzz dataset. Figure~\ref{Figure-every-bitmap} shows that \codename can achieve higher or at least the same bitmap size than other techniques on 14 out of 18 binaries.
For \emph{mujs}, \codename works worse than FairFuzz. For \emph{nm}, \codename works worse than EcoFuzz, AFLFast and AFL. For \emph{pdftotext}, \codename works worse than AFLFast. For \emph{tiffsplite}, \codename works worse than AFL and FairFuzz.

By comparing the performance of \codename with every baseline technique as shown in Figure~\ref{Figure-every-bitmap}, we can also observe that \codename outperforms \AFL, \AFLFast, \EcoFuzz, and \FairFuzz on 15, 15, 16, and 14 binaries, respectively.

As shown in Table~\ref{AFL++baseline-code-coverage}, \codenameplusplus achieves higher or at least the same bitmap size than AFL++ on 14 out of 18 binaries, AFL++-HIER on 15 out of 18 binaries.
\begin{table}
\small
        \begin{center}
        \setlength{\belowcaptionskip}{1.5mm}
        \caption{Averaged bitmap size and $p$ values on UniFuzz with \codenameplusplus as the baseline.}
        \vspace{-1.5em}
        \label{AFL++baseline-code-coverage}
        \resizebox{\columnwidth}{!}{
        \begin{tabular}{ l | c | c | c | c | c }
            \toprule
            \multirow{2}{1cm}{\bf{Binary}} & \multicolumn{1}{c|}{\bf{\codenameplusplus}} & \multicolumn{2}{c|}{\bf{AFL++}} & \multicolumn{2}{c}{\bf{AFL++-HIER}} \\
            \cline{2-6} & \bf{AVG cov} & \bf{AVG cov} & \bf{$p$ value} & \bf{AVG cov} & \bf{$p$ value}\\
            \midrule 
            cflow & \emph{\textbf{4.91\%}} & 4.69\% & \emph{\textbf{0.0008}} & 4.37\% & \emph{\textbf{\textless 0.0001}} \\
            exiv2 &\emph{\textbf{28.85\%}} & 24.23\%	& \emph{\textbf{0.0008}} & 	25.00\% &0.0635\\
            flvmeta & 2.15\%	&2.15\% &	\textgreater 0.9999 &	2.15\% & \textgreater 0.9999\\
            gdk & \emph{\textbf{10.44\%}} &	10.19\%&\emph{\textbf{0.00476}} &	9.26\% &\emph{\textbf{0.0016}}\\
            imginfo & 8.64\%	&8.65\%& 0.3155 &	8.70\% &\emph{\textbf{0.0476}}\\
            infotocap &\emph{\textbf{5.97\%}}&	5.78\%&	\emph{\textbf{0.0079}} &	2.64\% &\emph{\textbf{\textless 0.0001}}\\
            jhead &  1.39\%&	1.39\%&	\textgreater 0.9999 &	1.39\% &\textgreater 0.9999\\
            jq &  \emph{\textbf{8.51\%}}	&7.60\%&\emph{\textbf{0.0008}} &	7.68\% &\emph{\textbf{0.0079}}\\
            lame & \emph{\textbf{14.57\%}} &	14.33\%	&\emph{\textbf{0.0079}} &	14.01\% &\emph{\textbf{0.0023}}\\
            mp3gain & \emph{\textbf{4.13\%}} &	3.78\%	&\emph{\textbf{0.0008}}	&3.90\% & 0.3810\\
            mp42aac & 7.37\% &	6.89\%	&\emph{\textbf{0.0054}} 	&7.66\% &0.5238\\
            mujs & \emph{\textbf{13.57\%}} &	13.55\%	&\emph{\textbf{0.0412}} 	&13.45\% &\emph{\textbf{\textless 0.0001}}\\
            nm & 7.56\%	& 7.27\%	&\emph{\textbf{0.0079}} &	7.56\% &\emph{\textbf{0.0016}} \\
            objdump & \emph{\textbf{15.28\%}} &	13.57\%	&\emph{\textbf{\textless 0.0001}} &	13.82\% &\emph{\textbf{\textless 0.0001}}\\
            pdftotext & 28.13\%	&28.22\%&	0.5004 &	28.45\% &\emph{\textbf{0.0031}}\\
            tcpdump &  \emph{\textbf{30.41\%}}	&27.66\%	&\emph{\textbf{0.0008}} &	30.36\% &0.6962\\
            tiffsplit &6.20\%	&6.77\%	& 0.0519 &	5.77\% & \emph{\textbf{\textless 0.0001}}\\
            wav2swf & 1.29\%	&1.29\%	&\textgreater 0.9999 &	0.87\% & \emph{\textbf{\textless 0.0008}}\\
            \bottomrule
        \end{tabular}
        }
    \end{center}
\vspace{-1.5em}
\end{table}
 
\subsection{Statistical Significance}
To quantify whether there is a significant difference between \codename and other techniques, we leverage the Mann-Whitney U test to calculate the $p$ value~\cite{pvalue,Evaluating} using \codename as the baseline. According to Mann-Whitney U test~\cite{pvalue}, a $p$ value less than $0.05$ indicates a significant difference between the two fuzzers. 

As shown in Table~\ref{pvalue}, Column 2 shows the average code coverage of \codename. Column 3-6 show the $p$ values calculated by taking the code coverage of \codename as the baseline. We can observe that \codename significantly outperforms \AFL, \AFLFast, \EcoFuzz, and \FairFuzz on 7, 9, 8, and 10 binaries, respectively. 

Similarly, Table~\ref{AFL++baseline-code-coverage} shows the $p$ values calculated by taking the code coverage of \codenameplusplus as the baseline. We can observe that \codenameplusplus significantly outperforms AFL++ on 12 out of 18 binaries, and outperforms AFL++-HIER on 9 out of 18 binaries.

\subsection{Fuzzing throughput}
Our seed scheduling strategy introduces both positive and negative impacts on the fuzzing throughput. On the one hand, the tree structure can improve the fuzzing throughput in terms of time complexity. The searching time complexity of a tree structure is $O(log(N))$, whereas baseline fuzzing techniques organize the seed inputs in a queue, and the searching time complexity is $O(N)$. On the other hand, the implementation of our strategy requires collecting extra information, which may bring negative impacts on fuzzing throughput.
To investigate the overhead of our strategy, we compare the average number of executions per second by AFL++, \codenameplusplus, and AFL++-HIER on the UniFuzz dataset.

\begin{figure}
\centering
\includegraphics[width=0.9\columnwidth]{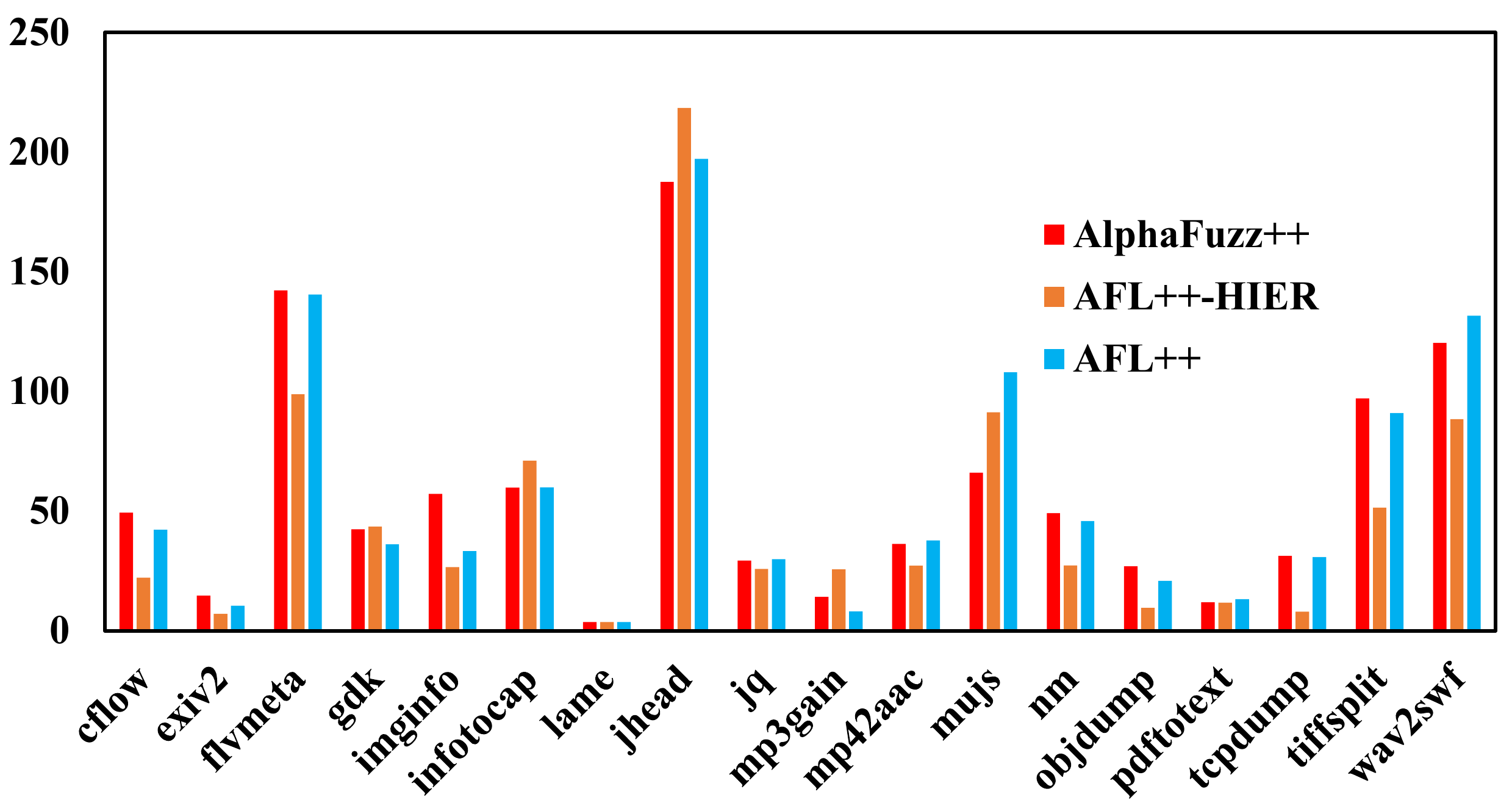}
\vspace{-1.0em}
\caption{Throughput comparison.}
\vspace{-2em}
\label{throughput}
\end{figure}

As shown in Figure~\ref{throughput}, \codenameplusplus's throughput is no less than \AFL on 12 out of 18 programs. That is, \codenameplusplus has a competitive throughput as AFL++.
In addition, \codenameplusplus has higher executions per second than AFL++-HIER on 11 out of 18 programs. 
The result verifies our analysis that AFL++-HIER requires maintaining multiple coverage metrics, which reduces the fuzzing's throughput. 

\subsection{Impact of the parameter k}\label{sec:parameter_k}
The parameter $k$ in \textbf{Formula~\ref{F_seed-score}} is a constant, which controls the trade-off between exploration and exploitation. According to the empirical value proposed by Kocsis and Szepesv$\acute{a}$ri~\cite{kocsis2006improved}, the value of $k$ is usually set to $\sqrt{2}$ or 1.414.

To investigate the impact of different $k$ values on seed scheduling, we assign $k$ as five different values $0$, $0.014$, $0.14$, $1.4$, and $14$, respectively. Then, we conduct experiments on two datasets, the CGC dataset and the UniFuzz dataset. 

Evaluation results show that the impact of $k$ varies a lot for different datasets. Specifically, as shown in Table~\ref{Table-parameter-k}, \codename achieves the highest code coverage with $k$ as 1.4 than other configurations on the CGC dataset. However, on the UniFuzz dataset, \codename achieves the highest code coverage when $k$ is set to 0.014. 

For in-depth analysis, we further examine the fuzzing processes on different datasets. Our manual analysis shows that the real-world binaries in UniFuzz are larger in scale and more complex than CGC binaries. Meanwhile, according to \textbf{Formula~\ref{F_seed-score}}, the value of $k\sqrt{\frac{lnN_i}{n_i}}$ for a seed will decrease with the increasing number of selected times. Thus, a smaller value of $k$ will contribute to exploitation, which lead ``seed mutation tree'' to expand in the depth direction. Therefore, we can infer that $k$ should be set as a smaller value with the increasing scale and more complex program structure.

\begin{table}[]
\centering
  \setlength{\belowcaptionskip}{1.5mm}
  \caption{The Averaged edge coverage on CGC and UniFuzz with different values of the parameter $k$.}
\begin{tabular}{l | c| c| c| c| c }
    \toprule
    \multirow{2}{1cm}{\bf{Dataset}} & \multicolumn{5}{c}{\bf{Value of $k$}}\\
            \cline{2-6} & \bf{0} & \bf{0.014} & \bf{0.14} & \bf{1.4} & \bf{14}\\
    \midrule
    CGC & 2.59\% & 2.61\% & 2.65\% & 2.75\% & 2.58\% \\
    UniFuzz & 8.38\% & 8.46\% & 8.38\% & 8.23\% & 8.22\% \\
    \bottomrule
\end{tabular}
\vspace{-1.5em}
\label{Table-parameter-k}
\end{table}

\subsection{Vulnerability detection on real-world binaries.}
To answer \textbf{RQ5}, we run the fuzzing techniques on 12 real-world binaries listed in Table~\ref{Table-linux}. We collect the crashes discovered by each fuzzing technique, and then leverage \emph{AddressSanitizer}~\cite{addresssanitizer} and \emph{GDB} to distinguish redundant crashes and identify unique vulnerabilities.
Table~\ref{Table-CVE} shows that these fuzzing techniques totally discover 12 vulnerabilities. We report the 12 vulnerabilities to upstream vendors. Among these vulnerabilities, 7 of them are confirmed by the vendors with the CVE-ID, 4 of them are confirmed but without a CVE-ID, and 1 of them has not been confirmed yet. Notably, we discovered 3 new vulnerabilities and obtained 3 \textbf{NEW CVEs}, which are \textbf{CVE-2021-25792}, \textbf{CVE-2021-25793}, and \textbf{CVE-2021-25794}. The POCs of the above vulnerabilities are available at (URL omitted for double-blind reviewing).

Table~\ref{Table-CVE} shows that \codename discovers more vulnerabilities than baseline techniques. Specifically, \codename discovers 5 vulnerabilities that are never detected by other techniques, and misses a vulnerability that is detected by \FairFuzz.

\begin{table}
  \centering
  \setlength{\belowcaptionskip}{1.5mm}
  \caption{Vulnerabilities discovered in real-world binaries.}
  \label{Table-CVE}
  \resizebox{\columnwidth}{!}{
  \begin{tabular}{c|c|c|c|c|c|c}
    \toprule
    \bf{Binary} & \bf{Vulnerabilities} & \bf{\AFL} & \bf{\AFLFast} & \bf{\EcoFuzz} & \bf{\FairFuzz} & \bf{\codename}\\
    \midrule
    \multirow{3}*{cjpeg} & CVE-2018-11214 & \checkmark  & \checkmark & \checkmark & \checkmark & \checkmark \\
    ~ & CVE-2018-11212 & \checkmark  & \checkmark & \checkmark & \checkmark & \checkmark \\
    ~ &  issue\textbf{\emph{\#5}} & \XSolidBrush & \XSolidBrush & \XSolidBrush & \XSolidBrush & \checkmark \\
    \hline
    \multirow{2}*{infotocap} & CVE-2018-19211 & \XSolidBrush  & \checkmark & \checkmark & \XSolidBrush & \checkmark \\
    ~ & \textbf{CVE-2021-25794}  & \XSolidBrush  & \XSolidBrush & \XSolidBrush & \XSolidBrush & \checkmark \\
    \hline
    \multirow{4}*{mp3gain} & CVE-2018-10778 & \checkmark & \checkmark & \checkmark & \checkmark & \checkmark \\
    ~ & CVE-2018-10777 &\checkmark & \checkmark & \checkmark & \checkmark & \checkmark  \\
    ~ & CVE-2017-14406 & \XSolidBrush  & \XSolidBrush & \XSolidBrush & \checkmark & \XSolidBrush\\
    ~ & CVE-2018-10776 & \XSolidBrush  & \XSolidBrush & \XSolidBrush & \XSolidBrush & \checkmark \\
    \hline
    \multirow{3}*{pdfimages} & issue\textbf{\emph{\#42073}} & \XSolidBrush & \XSolidBrush & \XSolidBrush & \XSolidBrush & \checkmark \\
    ~ & \textbf{CVE-2021-25792}  & \XSolidBrush & \XSolidBrush & \XSolidBrush & \XSolidBrush & \checkmark \\
    ~ & \textbf{CVE-2021-25793}  & \XSolidBrush & \checkmark & \XSolidBrush & \XSolidBrush & \checkmark \\
    \midrule
    \bf{Total} & \bf{12} & \bf{4} & \bf{6} & \bf{5} & \bf{5} & \bf{11} \\
    \bottomrule
  \end{tabular}
}
\end{table}
\section{Related work}
\myparagraph{Fuzzing.} In this paper, we mainly focus on seed scheduling for Coverage-based greybox fuzzing (CGF). 
Lots of fuzzing techniques leverage additional program analysis to improve the performance of fuzzing.
Honggfuzz~\cite{honggfuzz} and libfuzzer~\cite{serebryany2015libfuzzer} introduced a data flow feature—the degree of matching of the operands of branch statements, and prioritized the selection of seeds that more satisfies the branch constraints. GreyOne~\cite{greyone} uses data flow analysis to determine the relationship between input fields and constraint-related variables, then schedules the input with the highest number of relevant key fields. 

Besides CGF, lots of bug-driven seed scheduling strategies are proposed.
TortoiseFuzz~\cite{tortoiseFuzz} targets memory corruption vulnerabilities and prioritizes branches containing memory operations.
QTEP~\cite{qtep} prioritizes seeds that trigger more sensitive codes, so as to Dowser~\cite{Dowser}.
AFLgo~\cite{aflgo} prioritizes seeds that are closer to the target point to be tested. 

Besides the seed selection strategies, there are a lot of advanced fuzzing techniques focusing on seed generation~\cite{skyfire}, seed mutation~\cite{mopt}, coverage sensitivity~\cite{Jinghan}, and execution monitoring~\cite{Angora}.
MOPT~\cite{mopt} proposes a seed mutation strategy to determine the proper distribution for mutation operators.
NEUZZ~\cite{NEUZZ} identifies the significance of program smoothing and uses an incremental learning technique to guide the mutation of fuzzing.
Untracer~\cite{untracer} removes unnecessary instrumentation in basic blocks that have been explored to reduce the overhead of fuzzing.
Redqueen~\cite{redqueen} solves magic bytes and checksum tests via inferring input-to-state correspondence based on lightweight branch tracing.

\myparagraph{Monte Carlo Tree Search.} Monte Carlo Tree Search (MCTS) is a promising online planning approach~\cite{kocsis2006improved, browne2012survey, LiuWWGW20, BaierK20, Aleksander}.
The MCTS algorithm has achieved great success in Go~\cite{DeepMind} and has had a profound impact on the field of artificial intelligence~\cite{OuessaiSM20, DemediukT0R19, Bilal}.
The application field of MCTS algorithm is very extensive, such as active object recognition~\cite{patten2018monte}, wildlife monitoring\cite{hefferan2016adversarial}, environment exploration~\cite{efficient, planning}, and planetary exploration~\cite{arora2017approach}.
MCTS has been proposed in many different forms~\cite{browne2012survey}, but currently, the most common one is the Upper-Confidence Bounds Applied to Trees (UCT) algorithm\cite{kocsis2006improved}.
Therefore, we choose to use the UCT algorithm in our model.

MCTS algorithm also attracts the attention of researchers who focus on fuzzing. AFL-HIER~\cite{aflhier} proposes a multi-level coverage metric, and leverages UCT algorithm to perform seed selection. Legion~\cite{legion} leverages MCTS algorithm to maintain a balance between concolic execution and fuzzing, and improves the performance of hybrid fuzzing.

\section{Conclusion}
In this study, we make a key observation that the mutation relationships among seeds are valuable for seed scheduling. To investigate the seed mutation relationships, we design a ``seed mutation tree'' and further propose MCTS-based seed scheduling strategy by modeling the seed scheduling problem as a Monte-Carlo Tree Search (MCTS) problem. Evaluation shows that our approach outperforms other seed scheduling strategies with higher code coverage and more discovered vulnerabilities.

\bibliographystyle{plain}
\bibliography{mybibfile.bib}

\end{document}